\documentclass{ws-procs9x6}

\begin{document}

\title{$\Xi$ spectroscopy in photoproduction on a proton target at Jefferson Lab\footnote{\uppercase{T}his work is supported by \uppercase{DOE} under contract \uppercase{DE-AC05-84ER40150}}}

\author{L. Guo and D.~P. Weygand\\ For the CLAS Collaboration}
\address{Jefferson Lab, \\
12000 Jefferson Ave.,
Newport News, VA 23606, USA\\
E-mail: lguo@jlab.org}

\maketitle

\abstracts{
The CLAS Collaboration at Jefferson Lab conducted a photoproduction experiment on a proton target  using a tagged photon beam with an energy range of 1.5-3.8~GeV during May-July 2004. With an integrated luminosity of about 75 $pb^{-1}$, this experiment provides the largest data set for photon-proton reactions ever collected. The reaction $\gamma p \rightarrow K^+K^+\Xi^{-}(1320)$ has been investigated and the preliminary results of the cross section measurement of $\Xi^-(1320)$ for the photon energy range of 2.7-3.8~GeV have been obtained. In a search for excited cascade states, the reaction of $\gamma p \rightarrow K^+K^+\pi^-(\Xi^0(1320))$ has also been explored.
}

\section{Introduction and Photoproduction of cascade resonances}
Compared with the non-strange baryons and $S=-1$ hyperon states, the $\Xi$ resonances are in general not well studied. Only two cascade states, the $\Xi(1320)$ and $\Xi(1530)$, have four star status in the PDG. This is mainly due to small cross sections, and the fact that the cascade resonances cannot be produced through direct formation.
It is important to view the investigation of the cascade resonances as an essential part of  the baryon spectroscopy program.  Flavor $SU(3)$ symmetry implies as many $\Xi$ resonances as $N^*$ and $\Delta^*$ states, meaning many more cascade resonances await to be discovered~\cite{NEFKENS}. 
The facility at CEBAF offers the opportunity to study $\Xi$ resonances in exclusive photoproduction. Kaon beam experiments that dominated the earlier $\Xi$ studies suffer from low intensity, while hyperon beam experiments typically suffer from the high combinatoric background that complicates the analysis. It has been demonstrated that cascade production can be investigated through exclusive reactions such as $\gamma p \rightarrow K^+ K^+ (X)$~\cite{Price} in CLAS. However, low statistics hampered further investigation of the production mechanisms.

\section{Ground state $\Xi^-$ production in $\gamma p \rightarrow K^+K^+ (X)$}
\begin{figure}[htbp]
\begin{center}
\includegraphics[width=4in, height=1.3in]{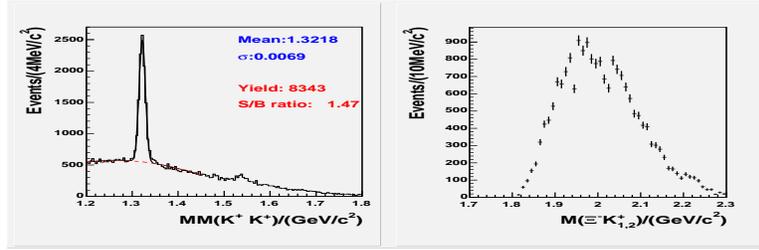}
\caption{Left: $K^+K^+$ missing mass spectrum with tighter timing cuts (less stringent cuts yield more than 12~k $\Xi^-$); Right: $\Xi^-K^+$ invariant mass spectra after side band subtraction (Both $K^+$ are included).}
\label{casm}
\end{center}
\end{figure}

\begin{figure}[htbp]
\begin{center}
\includegraphics[width=4in, height=1.8in]{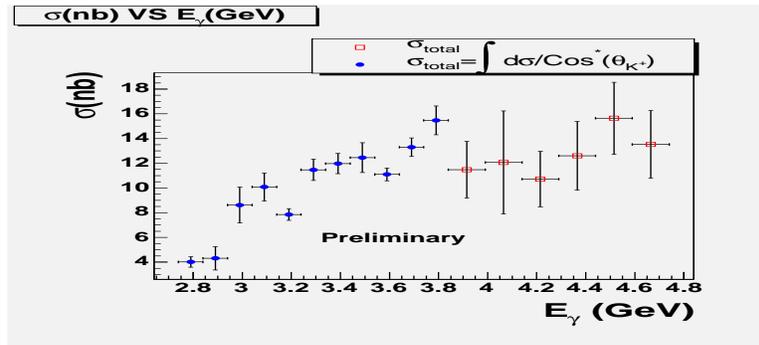}
\caption{Total cross sections(green) of $\Xi^-(1320)$. Results above 3.8~GeV are obtained without differential cross section measurement due to the very low statistics. Only statistical errors are shown.}
\label{total}
\end{center}
\end{figure}

The new CLAS photoproduction data set (g11) collected during summer 2004 used a tagged photon beam~\cite{CLAS}$^{,}$~\cite{TAGGER} mainly in the range of 1.5-3.8~GeV (less than 5\% data was collected with higher beam energy up to 4.75~GeV) incident on a proton target, and represents more than 75~pb$^{-1}$ luminosity. The ground state $\Xi^-$ is observed clearly in the $K^+K^+$ missing mass spectrum (Fig.~\ref{casm}, left), with the two kaons detected by the CLAS spectrometer. The first excited cascade resonance $\Xi^-(1530)$ is also visible. As discussed earlier, it is likely that the cascade resonances are produced through other intermediate resonances such as $Y^*$ states. However, there has not been unambiguous evidence for such a production mechanism in photoproduction. The $\Xi^-K^+$ invariant mass spectrum  (Fig.~\ref{casm}, right) seems to show interesting structures near 2~GeV. However, the two kaons being identical bosons and the uncertainty of the many overlapping hyperon states near 2~GeV make it difficult to determine the relevant parameters.
The $\Xi^-$ cross section has been extracted by assuming a t-channel process producing hyperon states that decay to $\Xi^-K^+$. Various differential cross section results have been studied to improve the simulation, and the total cross sections integrated from different methods are consistent with each other. It is worthwhile to note that the cross section\footnote{The cross sections are preliminary, and 20\% systematic errors are expected} of $\Xi^-$ seems to increase with the photon energy (Fig.~\ref{total}). This would be possible if the hypothesis of $Y^*\rightarrow \Xi^- K^+$ is correct since higher photon energy naturally leads to more accessible $Y^*$ states. The JLAB-MSU phenomenological approach~\cite{MOKEEV} for exclusive reactions with three final particles is being developed to incorporate the $\Xi^-K^+K^+$ channel and possibly determine the $\Xi^-$ photoproduction mechanism in the future.

\section{Reaction $\gamma p \rightarrow K^+K^+\pi^-(\Xi^0(1320))$}
\begin{figure}[htbp]
\begin{center}
\includegraphics[width=4in, height=2.8in]{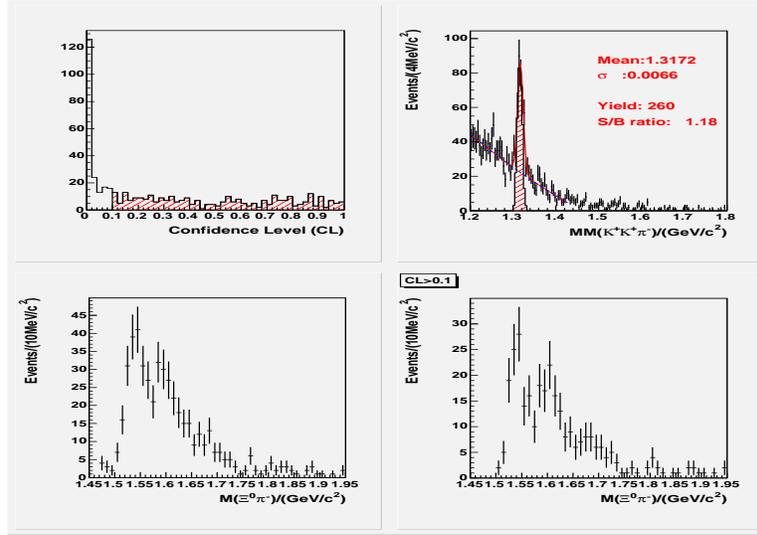}
\caption{Top Left: Confidence level (CL) distribution for events in the $\Xi^0$ peak (as shown on the right, shaded region indicates CL$>0.1$); Top right: ($K^+K^+\pi^-$) missing mass spectra (shaded region indicates CL$>0.1$); Bottom left: ($\Xi^0\pi^-$) invariant mass spectrum for events in the $\Xi^0$ peak; Bottom Right: ($\Xi^0\pi^-$) invariant mass spectrum (using kinematic fitted four vectors) for events in the $\Xi^0$ peak and CL$>0.1$; }
\label{fit4c}
\end{center}
\end{figure}

Excited cascade resonances can be investigated through the reaction $\gamma p \rightarrow K^+ K^+ \pi^- (\Xi^0)$ using the new CLAS data set. The $\Xi(1530)$ decays to $\Xi\pi$ almost $100\%$. For higher mass cascade resonaces, both $\Xi(1620)$ (one-star state) and $\Xi(1690)$ have been observed in the $\Xi\pi$ channel, but further confirmation is needed. The three charged particles are identified by the CLAS spectrometer, while the $\Xi^0$ events are reconstructed from missing mass (Fig.~\ref{fit4c}, top right). Kinematic fitting procedures were applied by constraining the $\Xi^0$ mass. The $\Xi^-(1530)$ peak is seen to become narrower after the fitting (Fig.~\ref{fit4c}, bottom), while a small enhancement around the 1605 MeV region is also observed. Due to the limited statistics, and the possible interference of $\gamma p \rightarrow K^+ K^{*0}(\Xi^0)$, it is difficult to draw definite conclusions regarding the origin of this small structure.

\section{Discussion}
More than 10000 $\Xi^-$ events have been observed in the reaction of $\gamma p \rightarrow K^+ K^+ (\Xi^-)$. The total cross section appears to increase with photon energy. Although the data suggest that the ground state $\Xi^-$ could be a decay product of high mass hyperons, the current status of the analysis is not conclusive. The reaction of  $\gamma p \rightarrow K^+ K^+ \pi^- (\Xi^0)$ has also been investigated. Although a small enhancement around 1605~MeV is observed near the $\Xi^-(1530)$ peak, the nature of this enhancement is not clear.
\section{Acknowledgment}
We wish to thank all of the CLAS collaborators, the extraordinary efforts of the CEBAF staff, and particularly the g11 running group.

\end{document}